# Designing for Different Stages in Behavior Change

Evangelos Karapanos

Cyprus University of Technology, Limassol, Cyprus
evangelos.karapanos@cut.ac.cy

**Abstract.** The behavior change process is a dynamic journey with different informational and motivational needs across its different stages; yet current technologies for behavior change are static. In our recent deployment of Habito, an activity tracking mobile app, we found individuals 'readiness' to behavior change (or the stage of behavior change they were in) to be a strong predictor of adoption. Individuals in the contemplation and preparation stages had an adoption rate of 56%, whereas individuals in precontemplation, action or maintenance stages had an adoption rate of only 20%. In this position paper we argue for behavior change technologies that are tailored to the different stages of behavior change.

## 1 Introduction

Despite their initial promise, physical activity trackers are failing to sustain users' engagement in the long run [1]. Shih et al. [2] found 50% of users who adopted a Fitbit to abandon it within the first two weeks of use. Similarly, we found [3] 62% of the users who downloaded an activity tracking mobile app to stop using it within the first two weeks, while in an online survey, one third of owners of activity trackers self-reported that they discarded them within six months after the purchase [4].

The question arises: is this a sign of activity trackers' failure to instill behavior change, or is this a positive sign in the sense that the tracker enabled the swift adoption of exercising by users as an intrinsically motivated practice, and exercising was no longer required (see [3])?

In a longitudinal field study of Habito [3], an activity tracking mobile app, we set to explore how individuals adopt and engage with activity trackers. Our study showed that things often do not go as we designers expect them to. For instance, contrary to conventional wisdom in the quantified-self community that behavior change is the result of deep knowledge about one's own behaviors, we found that people rarely look back at their past performance data and may not have deep knowledge about their own behaviors. Instead, we found the use of the tracker to be dominated by glances: brief, 5-sec sessions where users call the app to check how much they have walked so far without any further interaction. But activity trackers are not designed with glanceable interaction in mind.

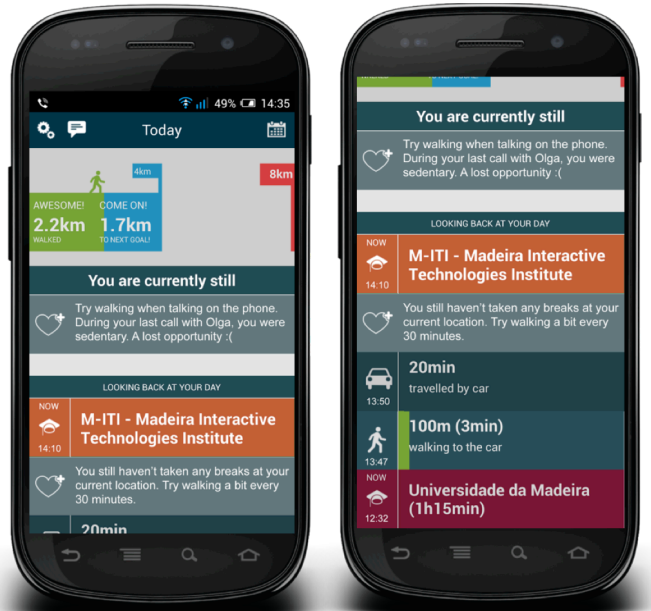

**Figure 1.** By deploying *Habito* [3] on Google Play we were able to monitor its usage over a 10-month period by 256 users who installed it on their own volition. *Habito* employs three design strategies to promote behavior change: *goal setting*, *contextualizing physical activity* and *textual feedback that keeps updating*.

Similarly, one of the most common design strategies in activity trackers is 'goal setting' - a user sets his or her own walking goal (e.g., 8 km per day) and feedback is provided as to how far he or she is from accomplishing the goal. But, while goal setting is a theoretically and empirically grounded strategy one could bring to design, it assumes that people self-set their own goals. Our study found that only 30% of users set their own goal, while 80% of users who did so, never updated the goal again (while updating one's goal would be expected in the process of behavior change).

Perhaps most interestingly, we found that current physical activity trackers work only for people that are in the intermediary stages of behavior change: those that have the motivation to change their behaviors but have no developed plans for doing so. Individuals in the contemplation and preparation stages, who have the intention but not yet the means (i.e. motivation, strategies) to change, had an adoption rate of 56% (with adoption being defined as use that extends beyond the first two weeks), whereas individuals in precontemplation, action or maintenance stages had an adoption rate of only 20%.

Yet, these individuals (in the intermediary stages of behavior change) are only about 43% of the population that are likely to purchase an activity tracker, or down-load an app on their smartphones (based on our sample [4]). So, there is a significant population of users for whom we currently fail to address their needs. To remediate this situation, we need to ask new questions, such as, how can trackers instill initial motivation for behavior change rather than merely supporting the process of it? Individuals in the precontemplation stage are often unaware of the extent of their inactivity [5]. As a result, initial experiences are marked by dismay as individuals realize their low activity levels. Rather than confronting users with this "truth", one could ask how trackers could increase individuals' perceptions of self-efficacy and competence and support them in the gradual increase of physical activity.

A second challenge is detecting the stage of behavior change individuals are in from behavioral cues. In doing so, one should bear into account that transitions across stages are not always unidirectional. Individuals often relapse to prior stages of behavior change. When this occurs, some individuals "feel like failures – embarrassed, ashamed and guilty" [6]. Detecting those transitions is as critical as detecting the stage an individual is currently in. Future work should thus embrace behavior change as a dynamic journey, should seek to understand the experiential side of behavior change, and to design strategies that support individuals across the full spectrum of their journey.